\documentclass[12pt,preprint]{aastex}
\usepackage{amsmath}

\usepackage{epsfig}
\usepackage{xcolor}

\usepackage{lineno}

\usepackage{mathrsfs}

\newcommand{\green}{f_{\rm G}}
\newcommand{\Green}{F_{\rm G}}
\newcommand{\tauy}{y}
\newcommand{\radw}{{\tilde r}}
\newcommand{\dimega}{\tilde \omega}
\newcommand{\sigmaT}{{\sigma_{_{\rm T}}}}
\newcommand{\msun}{M_\odot}

\begin{document}

\title{An Analytical Fourier-Transformation Model for the Production \break of
Hard and Soft X-Ray Time Lags in AGNs: Application to 1H~0707-495}
\author{
\text{David C. Baughman$^1$ and Peter A. Becker$^1$}\\
\text{$^1$Department of Physics and Astronomy, George Mason University,} \\
\text{Fairfax, VA 22030-4444; dbaughm2@masonlive.gmu.edu, pbecker@gmu.edu}
}
\date{\today}

\begin{abstract}
The variability of the X-ray emission from active galactic nuclei is often characterized using time lags observed between soft and hard energy bands in the detector. The time lags are usually computed using the complex cross spectrum, which is based on the Fourier transforms of the hard and soft time series data. It has been noted that some active galactic nuclei display soft X-ray time lags, in addition to the more ubiquitous hard lags. Hard time lags are thought to be produced via propagating fluctuations, spatial reverberation, or via the thermal Comptonization of soft seed photons injected into a hot electron cloud. The physical origin of the soft lags has been a subject of debate over the last decade. Currently, the reverberation interpretation is recognized as a leading theory. In this paper, we explore the alternative possibility that the soft X-ray time lags result partially from the thermal and bulk Comptonization of monochromatic seed photons, which in the case of the narrow-line Seyfert 1 galaxy 1H 0707-495, may correlate with fluorescence of iron L-line emission. In our model, the seed photons are injected into a hot, quasi-spherical corona in the inner region of the accretion flow. We develop an exact, time-dependent analytical model for the thermal and bulk Comptonization of the seed photons based on a Fourier-transformed radiation transport equation, and we demonstrate that the model successfully reproduces both the hard and soft time lags observed from 1H 0707-495.

\smallskip

\textit{key words}: black hole physics - stars: black holes - X-rays: general - X-rays: individual (1H~0707-495)
\end{abstract}

\section{Introduction}

It has often been noted that all black hole sources demonstrate similar patterns of X-ray variability, with the nature of the variability dependent on whether the source is in the low/hard state or the high/soft state (see van der Klis 2006 for a review). The same general variability patterns are observed for black holes of all mass scales, ranging from Galactic black holes (GBHs) to supermassive active galactic nuclei (AGNs). For example, by investigating the power spectral densities (PSDs) for a large sample of GBHs and AGNs, McHardy et al. (2006) showed that the timescale, $t_{\rm break}$, associated with the location of the characteristic PSD break frequency scales as $t_{\rm break} \propto (M)(\dot M / \dot M_{\rm E})^{-1}$, where $M$ is the black hole mass, $\dot M$ is the accretion rate, and $\dot M_{\rm E} $ denotes the Eddington accretion rate. A similar mass scaling law was discovered by De Marco et al. (2013), who observed that the Fourier frequency associated with the onset of negative (soft) time lags, $\nu_{\rm neg}$, scales as $\nu_{\rm neg} \propto (\dot M / \dot M_{\rm E})/(M)$. These correlations suggest the universality of the underlying accretion process over many decades of black hole mass.

In this paper we investigate the structure of black hole accretion flows using studies of variability based on observations of Fourier X-ray time lags in AGN. Our results successfully replicate an important, and relatively new observation, while simultaneously demonstrating a similar mass scaling relationship to that described by De Marco et al. (2013).
In our model, the observed mass scaling laws occur as a natural consequence of bulk and thermal Comptonization occurring near the black hole.

\subsection{Fourier Time Lags}

The idea of computing the time lags between two photon energy bands using Fourier analysis was first suggested by van der Klis et al. (1987), with subsequent development and application by Nowak et al. (1999). The calculation of the time lags is based on the development of the complex cross spectrum, $C$, defined by
\begin{equation}
C(\omega) = S^{*}(\omega)H(\omega) \ ,
\label{eq:crossspectrum}
\end{equation}
where
\begin{equation}
\omega = 2 \pi \nu_F \ ,
\label{eq:omegadef}
\end{equation}
denotes the circular Fourier frequency, $\nu_F$ is the Fourier frequency in Hz, and $S(\omega)$ and $H(\omega)$ represent the Fourier transforms of the soft and hard photon energy light curves, corresponding to photon energies $\epsilon_s$ and $\epsilon_h$, respectively. The asterisk in Equation~(\ref{eq:crossspectrum}) represents the complex conjugate.

The Fourier transforms $S(\omega)$ and $H(\omega)$ appearing in Equation~(\ref{eq:crossspectrum}) are computed using the integrals
\begin{equation}
\begin{split}
S(\omega) &= \int_{-\infty}^{+\infty} s(t) \, e^{i \omega t} dt \ , \\
H(\omega) &= \int_{-\infty}^{+\infty} h(t) \, e^{i \omega t} dt \ ,
\end{split}
\label{eq:transforms}
\end{equation}
where the soft and hard photon energy light curves are denoted by $s(t)$ and $h(t)$, respectively. The two light curves are related to the time-dependent X-ray photon count-rate spectrum observed at the detector, $\mathscr F_\epsilon(\epsilon,t)$, via
\begin{equation}
\begin{split}
& s(t) = \mathscr F_\epsilon(\epsilon_s,t) \ , \\
& h(t) = \mathscr F_\epsilon(\epsilon_h,t) \ .
\end{split}
\label{eq:transforms2}
\end{equation}
The corresponding Fourier time lag, $\delta t$, is then evaluated using the relationship
\begin{equation}
\delta t(\omega) = \frac{\phi(\omega)}{\omega}  \ ,
\label{eq:lag}
\end{equation}
where the phase lag, $\phi$, is given by
\begin{equation}
\phi(\omega) = \rm{Arg}[S^{*}(\omega)H(\omega)]  \ .
\label{eq:plag}
\end{equation}

According to the sign convention adopted here, a positive time lag $\delta t > 0$ corresponds to a hard lag, in which the hard channel signal lags the soft signal for a given Fourier frequency (or inverse variability period). It can be shown that if an exact time delay is introduced between the hard and soft channel light curves $h(t)$ and $s(t)$ so that $h(t)=s(t-\Delta t)$, then the resulting time lag is $\delta(\omega)=\Delta t$ as expected.

It can also be easily deduced that the existence of time lags necessarily implies time variability in the source, since steady-state light curves would generate no phase lag, and therefore no time lag. Hence the detection of time lags from a given source necessarily implies the existence of variability in the X-ray signal from the source. This implies that the emission from a given source can be viewed as a superposition of continual steady-state emission, combined with episodic bursts or flashes, which produce time-dependent phenomenon such as the time lags. That is the approach we take here in our analysis of the emission from the narrow-line Seyfert 1 (NLS1) galaxy 1H~0707-495.

\subsection{Time Lag Observations}

Hard Fourier time lags have been detected in the X-ray emission from many AGNs over the past three decades, as discussed by Hasinger et al. (1986), van der Klis et al. (1987), Lyubarskii (1997), Crary et al. (1998), B\"ottcher \& Liang (1999), Kotov et al. (2001), Ar\'evalo \& Uttley (2006), and Kroon \& Becker (2016). However, a relatively new field of investigation of AGN X-ray variability involves the robust detection of {\it soft lags} from AGNs. First reported for  1H~0707-495 (Fabian et al. 2009), broader studies have revealed that soft lags are not uncommon, and are found in all AGN mass scales, ranging from low-mass (Mallick et al. 2021) to high-mass (De Marco et al. 2013). For example, De Marco et al. (2013) detected soft time lags in 15 of the 32 radio-quiet AGNs in their survey, and also determined to high accuracy $(\gtrsim 4\sigma)$ that the magnitude of the lags scaled in proportion to the black hole mass over 2.5 orders of magnitude. The universality of the scaling law suggests that the same underlying physical mechanism may be responsible for producing the soft lags in all AGNs, regardless of the black hole mass.

The time scales of the observed high-mass AGN soft lags are on the order of $\sim 10-100$~seconds (e.g., Fabian et al. 2009; Emmanoulopoulos et al. 2011; Cackett et al. 2013; De Marco et al. 2013), which are on the order of multiples of gravitational radii of AGN, making them extremely useful probes for understanding the structure of the inner regions of the accretion flows, including temperature, density, and accretion configuration. It should be noted that soft Fourier time lags have also been discovered in stellar-mass sources (e.g., De Marco et al. 2015). However, there is a great deal of difficulty in using these sources to study the underlying physical processes due to the extremely small time scales.

\subsection{Previous Models}

To date, two primary scenarios have emerged regarding the mechanism responsible for producing the soft Fourier time lags observed from GBHs and AGNs. The first scenario centers on the reprocessing of radiation in a zone in which the electrons cool on a timescale comparable to the lag time. The second scenario, commonly referred to as the reverberation interpretation, has become somewhat of a consensus in the literature, and focuses on the reflection of hard emission produced near the black hole off a cool electron population located at a larger radius, causing down-scattering with a time lag created by the transit-time propagation delay.

Malzac \& Jourdain (2000) investigated the first scenario, in which emission produced in the corona is down-scattered over time by electrons in the corona that cool in response to the scattering of the radiation. They predicted that if the flares arise from this mechanism, then the lags should invert from hard to soft if the energy dissipation timescale is between $10\,H/c$ and $100\,H/c$, where $H$ is the size of the emission region. These authors presented illustrative numerical simulations, but made no attempt to fit the data for any source.

When soft time lags were first discovered in the X-ray emission from 1H~0707-495, Zoghbi et al. (2010) argued that the model of Malzac \& Jourdain (2000), in which the spectrum is formed in a cooling corona, is inconsistent with the observed spectral variability around $\sim$ 1 keV. They proposed an alternative scenario in which the time lags are the result of spatial reverberation. In their model, the soft lags are formed when hard photons from the corona (generated near the event horizon) are reflected off the surrounding, cooler accretion disk, which extends from an inner radius of $\sim 1.3\,R_g$ to an outer radius $\sim 400\,R_g$, where $R_g=GM/c^2$ is the gravitational radius for the central black hole. Zoghbi et al. (2010) concluded that this type of reflection scenario naturally produces soft time lags with a magnitude comparable to the light travel time between the corona and the reflecting region of the disk. A time lag of $\sim 30\,$s would therefore correlate with a black hole mass of $M \sim 2 \times 10^6\,\msun$, which is consistent with the uncertain mass of this black hole (e.g., Zhou \& Wang 2005).

Miller et al. (2010) analyzed relative lags between three energy bands and concluded that their results invalidated the reflection model of Zoghbi et al. (2010). As an alternative, they proposed that the soft lags in 1H~0707-495 are caused by a partially opaque reverberating region with a minimum radius of $\sim 1,000$ light-seconds from the black hole corresponding to $\sim 100\,R_g$ for a black hole with mass $M \sim 2 \times 10^6\,\msun$. In Miller's model, this region has decreasing opacity with increasing energy and has scattering or reflection present in all X-ray bands. Zoghbi et al. (2011) argue that Miller's results are invalid because they imposed artificial constraints on the model geometry.

Subsequent authors have further explored the application of the reverberation model to the interpretation of the observed soft lags. Gardner \& Done (2014) investigated the NLS1 AGN PG1244+026 and proposed that reflection alone is unable to produce the soft lags without additional emission from the accretion flow as well as additional reprocessed emission. Emmanoulopoulos et al. (2016) used a method proposed by Papadakis et al. (2016) to search for evidence of reverberation lags in the PSD of 1H~0707-495. We note that the non-detection of interference fringes in the PSD was consistent with the model, since the observations had insufficient signal-to-noise, coupled with too much uncertainty regarding the shape of the underlying PSD of the direct emission, which made it impossible to rule out the null hypothesis. There have also been numerous studies examining the iron K features in the lag versus energy spectrum for AGNs and GBHs that provides supporting evidence for the reverberation lag hypothesis (e.g., Kara et al. 2016; Kara et al. 2019; Wang et al. 2021).

\subsection{Fourier Transform Model}

The ongoing controversy regarding the origin and interpretation of the soft time lags, as well as the extreme complexity of the existing theoretical models, has motivated us to develop a new model for the production of the observed hard and soft time lags observed from AGN. In this paper, we explore the possibility that the observed time lags may be the result of thermal and bulk Comptonization occurring in the freely-falling, quasi-spherical inner region (corona) of accretion flows onto supermassive black holes. In this scenario, the time-dependent part of the signal (that produces the observed time lags) is the result of the episodic emission of monochromatic photons generated near the black hole, which are then Comptonized by hot electrons in the quasi-spherical coronal region. Conversely, the steady-state X-ray spectrum in this model represents emission produced continuously by the optically thick, geometrically thin accretion disk. The episodic emission of the seed photons may be due to inhomogeneous ``clumps" in the accretion flow (e.g., Merloni et al. 2006; Guti\'errez et al. 2021).

The theoretical approach adopted here is a novel one, based on the fact that the observed time lags are computed using Fourier transformed X-ray light curves from the source AGN at two different X-ray energies (Kroon \& Becker 2016). This has motivated us to develop theoretical predictions for the time lags by directly solving the transport equation to determine the Fourier transform of the radiation distribution throughout the quasi-spherical coronal region near the center of the accretion flow. This method is convenient because it facilitates the direct computation of Fourier-based time lags, which can then be compared with observational data.

The time-dependent transport equation we solve includes terms describing bulk and thermal Comptonization, spatial diffusion, advection, and photon injection and escape. We demonstrate that the results obtained for the time lags as a function of Fourier frequency agree reasonably well with the observations for 1H~0707-495, without requiring reflection off the accretion disk, or any special geometrical constraints.

The remainder of the paper is organized as follows. In Section~\ref{sec:method}, we introduce the transport equation and discuss the dynamical model of the spherical accretion flow. In Section~\ref{sec:trans}, we obtain the exact solution for the Fourier transform of the radiation distribution inside the spherical flow. In Section~\ref{sec:compute} we develop the formalism for computing the X-ray time lags based on the Fourier transform of the radiation distribution escaping from the outer boundary of the spherical flow. In Section~\ref{sec:applications} we use the new model to generate simulated time lags for the narrow-line Seyfert 1 galaxy 1H~0707-495, and we compare the simulated lags with the observational data. Finally, in Section~\ref{sec:conclusions}, we provide an overview of our main results and their astrophysical significance. 


\section{Model Overview}
\label{sec:method}

In this paper, we explore an alternative model for the production of the observed hard and soft time lags from accreting black holes. The previously explored reverberation and reflection scenarios provide a natural mechanism for generating time lags via light transit-time effects, in which the reverberation is a geometrical effect due to the propagation and scattering of photons through the physical space of radius. On the other hand, the new model developed here explores the complementary effect of thermal and bulk Comptonization in creating time lags due to reverberation in the {\it energy space} rather than the physical space, as in the previous models. We argue that Comptonization lags and spatial transit-time lags probably both contribute to the observed time lags, and there is an interesting duality between the two mechanisms, since the Comptonization occurs in the energy space and geometrical reverberation occurs in the physical space.

The model developed here involves the analytical solution of a time-dependent transport equation describing the scattering of seed photons injected into an accreting plasma cloud in the region of flow near the central black hole. The seed photons are assumed to be injected as an instantaneous flash of energy driven by a monochromatic photon source.

Following the work of Colpi (1988), the transport equation considered here includes the effects of both bulk and thermal Comptonization, as well as the spatial transport of radiation via diffusion and advection. However, we extend the model of Colpi in two important ways. First, we investigate time-dependent solutions using a Fourier transform technique. Second, we impose free-streaming boundary conditions at finite values for the inner and outer radii of the region.

In the case of 1H~0707-495, the steady-state X-ray continuum spectrum is likely produced by an accretion disk, which may extend to an outer radius $\sim 10^3\,R_g$ (Zoghbi et al. 2010). Close to the black hole, the disk may extend to the radius of marginal stability, or, alternatively, it may transition into a freely-falling, quasi-spherical corona, either due to passage through a shock (e.g., Chakrabarti 1989; Das et al. 2009; Becker et al. 2011; Chattopadhyay \& Kumar 2016), or due to the accretion of a separate population of matter that is supplied with low angular momentum (e.g., Proga \& Begelman 2003; Moscibrodzka et al. 2007), or possibly due the removal of a substantial fraction of the angular momentum via a mildly relativistic wind (e.g., Dauser et al. 2012; Done \& Jin 2016). In this work, we assume that the disk either extends inward to the radius of marginal stability, with an overlying corona, or that the disk truncates near the outer radius of the corona due to one of the mechanisms listed above. Hence we assume that the time-dependent Comptonization is occurring in a quasi-spherical inner region, where the radial velocity, $v_r$, and the electron number density, $n_e$, follow the free-fall profiles $v_r \propto r^{-1/2}$ and $n_e(r)\propto r^{-3/2}$, respectively.

We assume that the electrons in the quasi-spherical inner region comprise a Maxwellian distribution with temperature $T_e$. This region is expected to be roughly isothermal due to the ``thermostat'' effect of thermal Comptonization (e.g., Sunyaev \& Titarchuk 1980). The isothermal assumption is also supported by the simulations of Meyer-Hofmeister et al. (2012) who show that thermal Comptonization leads to a weak radial dependence for the electron temperature in the corona. A simplified schematic representation of our model is provided in Figure 1.
\begin{figure}[h]
\centerline{\includegraphics[scale=0.4]{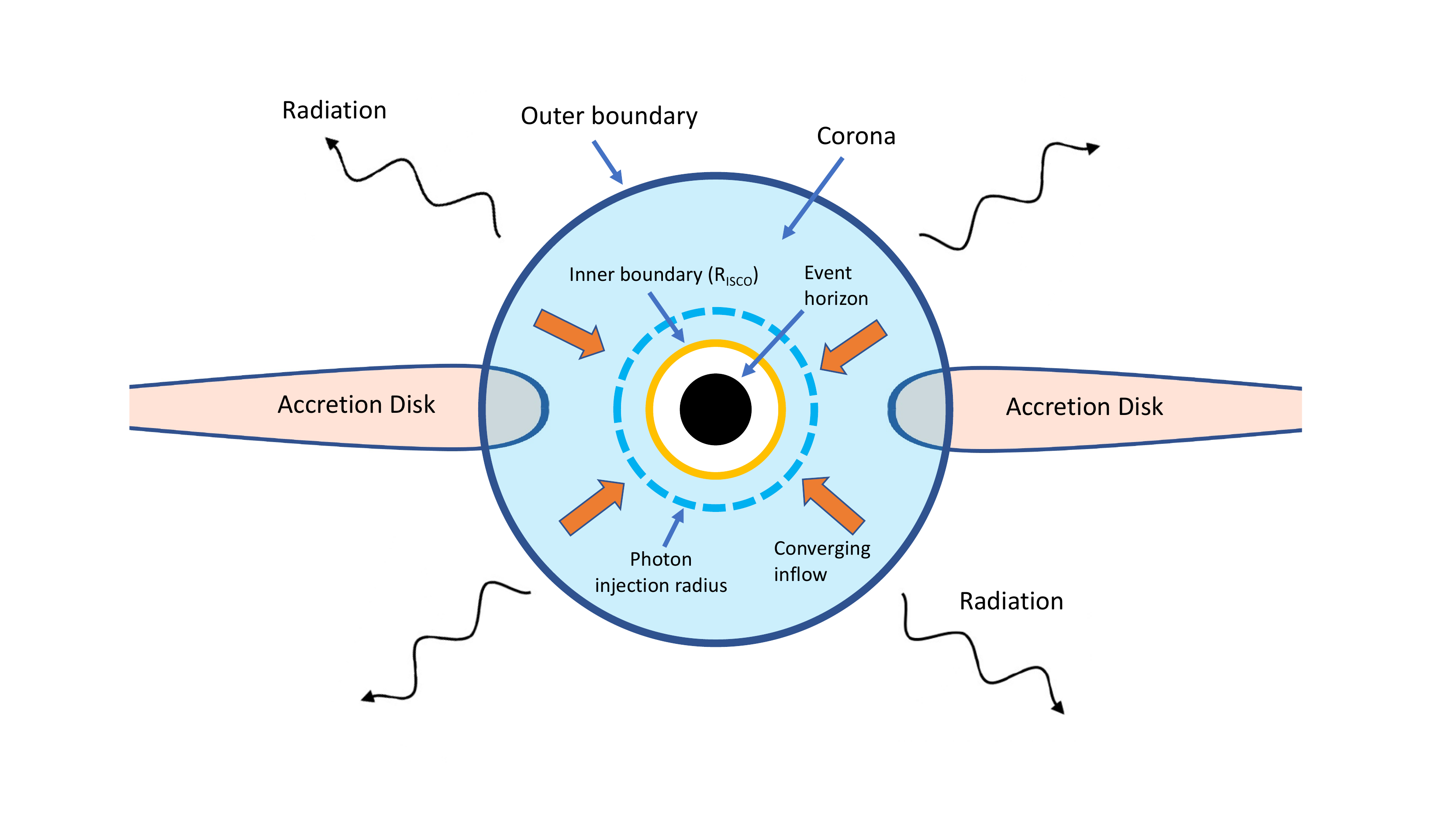}}
\vskip-0.4truein
\caption{\label{fig:AGN1}\footnotesize{The geometry of our physical model. A hot, inner coronal region is surrounded by a relatively cool accretion disk. The disk may truncate at the radius of marginal stability, or it may transition into the corona due to passage through a shock, or due to the removal of angular momentum via a mildly relativistic wind. See the discussion in the text.
}}
\end{figure}
%


\vspace{\baselineskip}

\subsection{Transport Equation}
\label{sec:transport}

Our radiation transport model expands upon the work of Colpi (1988) and Kroon \& Becker (2016) by incorporating the effects of time-dependent bulk and thermal Comptonization in a spherically-symmetric converging flow. We begin by writing down the general time-dependent radiation transport equation describing the evolution of the photon distribution function, $f(\vec r,\epsilon,t)$, in an arbitrary geometry, which is given by (Becker 2003)
\begin{equation}
\begin{split}
\frac{\partial f}{\partial t} = -\vec v\cdot\vec \nabla f + \vec \nabla\cdot\left(\kappa\vec \nabla f\right)
+ \left(\vec \nabla\cdot\vec v\right)\frac{\epsilon}{3}\frac{\partial f}{\partial \epsilon} \\
+ \frac{n_e \sigmaT c}{m_e c^2}\frac{1}{\epsilon^2}\frac{\partial}{\partial\epsilon}
\left[\epsilon^4 \! \left(f + kT_e\frac{\partial f}{\partial\epsilon}\right)\right] + Q \ ,
\label{eq:transport}
\end{split}
\end{equation}
where $\epsilon$ denotes the photon energy, $\vec r$ is the spatial location, $\vec v$ is the accretion velocity, $\sigmaT$ denotes the Thomson cross section, $\kappa$ is the spatial diffusion coefficient, $k$ denotes the Boltzmann constant, $T_e$ is the (constant) electron temperature, $m_e$ is the electron mass, $c$ is the speed of light, and $Q$ denotes the photon source term. The third and fourth terms on the right-hand side represent the effects of bulk and thermal Comptonization, respectively.

The radiation distribution function, $f$, is normalized so that the photon number and energy densities, $n_r$ and $U_r$, respectively, are computed using
\begin{equation}
n_r(\vec r,t) = \int_0^\infty \epsilon^2 f(\vec r,\epsilon,t) \, d\epsilon \ \ \propto ~{\rm cm}^{-3} \ ,
\label{eq:nden}
\end{equation}
and
\begin{equation}
U_r(\vec r,t) = \int_0^\infty \epsilon^3  f(\vec r,\epsilon,t) \, d\epsilon \ \ \propto~ {\rm erg \ cm}^{-3} \ .
\label{eq:enden}
\end{equation}

In a spherically-symmetric coronal region, the transport equation can be rewritten as
\begin{equation}
\begin{split}
\frac{\partial \green}{\partial t} = - v_r \frac{\partial \green}{\partial r} &+ \frac{1}{r^2}\frac{\partial}{\partial r}\left(r^2\kappa\frac{\partial \green}{\partial r} \right) + \frac{1}{r^2}\frac{\partial}{\partial r}\left(r^2 v_r\right)\frac{\epsilon}{3}\frac{\partial \green}{\partial \epsilon} \\
&+ \frac{n_e \sigmaT c}{m_e c^2}\frac{1}{\epsilon^2}\frac{\partial}{\partial\epsilon}
\left[\epsilon^4 \! \left(\green + kT_e\frac{\partial \green}{\partial\epsilon}\right)\right] \\
&+ \frac{N_0\delta(r-r_0)\delta(\epsilon-\epsilon_0)\delta(t-t_0)}
{4\pi r_0^2 \epsilon_0^2} \ ,
\end{split}
\label{eq:spheretransport}
\end{equation}
where $v_r < 0$ is the radial accretion velocity and $\green(r,r_0,\epsilon,\epsilon_0,t)$ denotes the Green's function, which is the response to the instantaneous injection of $N_0$ photons of energy $\epsilon_0$ at time $t_0$ from a source at radius $r_0$.

The fundamental transport equation (Equation~(\ref{eq:spheretransport})) is linear, and therefore it follows that once the solution for the Green's function, $\green$, has been obtained, the corresponding particular solution, $f$, associated with the general photon source function, $Q$, is given by the integral convolution
\begin{equation}
f(r,\epsilon,t) = \int_0^\infty \int_0^\infty \int_0^\infty \frac{\green}{N_0} \, \epsilon_0^2 \, Q \, 4 \pi r_0^2 \, dr_0 \, d\epsilon_0 \, dt_0
\ .
\label{eq:convolve}
\end{equation}
In our model, the bounds for the radial integration in Equation~(\ref{eq:convolve}) are replaced with the inner and outer radii of the quasi-spherical corona.

\subsection{Dynamics}
\label{sec:dynamics}

In the quasi-spherical coronal region under consideration here, the radial velocity, $v_r$, and the electron number density, $n_e$, are assumed to follow the free-fall profiles $v_r \propto r^{-1/2}$ and $n_e(r)\propto r^{-3/2}$, respectively. The radial velocity $v_r(r)$ profile can therefore be written as
\begin{equation}
v_r(r) = \ell \, v_{\rm ff}(r) = - \ell \, \sqrt{\frac{2GM}{r}} \ ,
\label{eq:v}
\end{equation}
where $v_{\rm ff}(r)$ denotes the local free-fall velocity, and the constant $\ell$ lies in the range $0 \le \ell \le 1$, with $\ell=1$ corresponding to exact free-fall (Colpi 1988).

Assuming that the accreting gas is composed of pure, fully-ionized hydrogen, the spatial diffusion coefficient, $\kappa$, appearing in Equation~(\ref{eq:spheretransport}) is related to the electron number density, $n_e$,
via
\begin{equation}
\kappa(r) = \frac{c}{3 n_e(r) \sigmaT} \ .
\label{eq:kappa}
\end{equation}
The electron number density, $n_e$, is related to the mass accretion rate, $\dot M$, via
\begin{equation}
\dot M = 4 \pi r^2 m_p n_e |v_r| = \text{constant} \ ,
\label{eq:bigmdot}
\end{equation}
where $m_p$ denotes the proton mass. Combining Equations~(\ref{eq:v}), (\ref{eq:kappa}), and (\ref{eq:bigmdot}), we note that the radial velocity, $v_r$, the electron number density, $n_e$, and the spatial diffusion coefficient, $\kappa$, can be written as
\begin{equation}
v_r(\radw) = - \hat v c \, \radw^{-1/2} \ , \ 
n_e(\radw) = \frac{\radw^{-3/2}}{3 \, \sigmaT R_g \hat\kappa} \ , \ 
\kappa(\radw) = \hat\kappa R_g c \, \radw^{3/2} \ ,
\label{eq:nondeq1}
\end{equation}
where $R_g = GM/c^2$, and we have introduced the dimensionless constants $\hat v$ and $\hat\kappa$ and the dimensionless radius
\begin{equation}
\radw \equiv \frac{r}{R_g} \ .
\label{eq:dimrad}
\end{equation}

We note that the constant $\hat v$ introduced in Equation~(\ref{eq:nondeq1}) is related to Colpi's constant $\ell$ via $\hat v = \sqrt{2} \, \ell$. Hence setting $\hat v=\sqrt{2}$ yields the exact free-fall case with $v_r(r) = v_{\rm ff}(r)$ (see Equation~(\ref{eq:v})). We also note that Equations~(\ref{eq:bigmdot}) and (\ref{eq:nondeq1}) can be combined to relate $\hat v$ and $\hat \kappa$ to the accretion rate $\dot M$ via
\begin{equation}
\dot M = \frac{4 \pi R_g m_p \hat v c}{3 \hat\kappa \sigmaT} \ .
\label{eq:MdotNew}
\end{equation}

Transforming the spatial coordinate from $r$ to $\radw$ in the transport equation and substituting for $v_r$, $n_e$, and $\kappa$ using Equation~(\ref{eq:nondeq1}) yields
\begin{equation}
\begin{split}
\frac{\partial \green}{\partial t} &= \frac{\hat v c \radw^{-1/2}}{R_g} \frac{\partial \green}{\partial \radw}
+ \frac{\hat \kappa c \radw^{-2}}{R_g} \frac{\partial}{\partial \radw} \left(\radw^{7/2} \frac{\partial \green}{\partial \radw}\right) \\
&- \frac{\hat v c \radw^{-3/2}}{2 R_g} \chi\frac{\partial \green}{\partial \chi} \\
&+ \frac{c \Theta \radw^{-3/2}}{3 \hat \kappa R_g}
\frac{1}{\chi^2}\frac{\partial}{\partial \chi}
\left[\chi^4 \! \left(\green + \frac{\partial \green}{\partial \chi}\right)\right] \\
&+ \frac{N_0\delta(\radw-\radw_0)\delta(\chi-\chi_0)\delta(t-t_0)}
{4\pi r_0^2 \epsilon_0^2 k T_e R_g} \ ,
\end{split}
\label{eq:transeqw1}
\end{equation}
where we have also introduced the dimensionless photon energy, $\chi$, and the dimensionless temperature, $\Theta$, using
\begin{equation}
\begin{split}
\chi \equiv \frac{\epsilon}{kT_e} \ , \quad \Theta \equiv \frac{kT_e}{m_e c^2} \ .
\end{split}
\label{eq:nondeq2}
\end{equation}

\subsection{Trapping Radius}
\label{sec:traprad}

In spherically-symmetric accretion flows, the spatial transport of the radiation is dominated by inward-bound advection for $r < r_t$, and by outward-bound diffusion for $r > r_t$, where $r_t$ is the ``trapping radius,'' defined by (Payne \& Blandford 1981; Colpi 1988)
\begin{equation}
r_t \equiv \frac{3 \dot M \sigmaT}{4 \pi m_p c} \ .
\label{eq:traprad}
\end{equation}
We can also use Equation~(\ref{eq:MdotNew}) to express the trapping radius in the form
\begin{equation}
\frac{r_t}{R_g} = \frac{\hat v}{\hat \kappa} = 3 \dot m \ ,
\label{eq:traprad2}
\end{equation}
where the dimensionless accretion rate, $\dot m$, is defined by
\begin{equation}
\dot m \equiv \frac{\dot M}{\dot M_{\rm E}} \ ,
\label{eq:mdot}
\end{equation}
and the Eddington accretion rate, $\dot M_{\rm E}$, is defined by
\begin{equation}
\dot M_{\rm E} \equiv \frac{4 \pi G M m_p}{c \sigmaT} \ .
\label{eq:eddMdot}
\end{equation}

It is also of interest to compute the electron scattering optical depth, $\tau(r)$, between radius $r$ and the outer radius of the spherical coronal region, located at $r=r_{\rm out}$. The result obtained is
\begin{equation}
\tau(\radw) = \int _r^{r_{\rm out}} n_e(r') \sigmaT \, dr'
= \frac{2 \, \dot m}{\hat v} \left(\frac{1}{\sqrt{\radw}} - \frac{1}{\sqrt{\radw_{\rm out}}}\right) \ ,
\label{eq:tau}
\end{equation}
where $\tilde r = r/R_g$, $\tilde r_{\rm out} = r_{\rm out}/R_g$, and we have utilized Equation~(\ref{eq:nondeq1}) to substitute for $n_e(r)$.

As noted by Payne \& Blandford (1981) and Colpi (1988), the physical significance of the trapping radius motivates a transformation of the spatial variable $\radw$ in the transport equation to the new spatial variable $\tauy$, defined by
\begin{equation}
\tauy \equiv \frac{r_t}{r} = \frac{\hat v}{\hat \kappa\radw} \ ,
\label{eq:traprad1}
\end{equation}
so that the trapping radius corresponds to $\tauy = 1$, and the radiation is trapped in the region with $\tauy > 1$ ($r < r_t$). Making the change of variables in the transport equation from $\radw$ to $\tauy$ yields
\begin{equation}
\begin{split}
\frac{\partial \green}{\partial q} &= \frac{\Theta}{3\hat\kappa \hat v } \frac{\tauy^{3/2}}{\chi^2}\frac{\partial}{\partial \chi}
\left[\chi^4 \! \left(\green + \frac{\partial \green}{\partial \chi}\right)\right]  \\
& + \tauy^{4} \frac{\partial}{\partial \tauy} \left(\tauy^{-3/2}  \frac{\partial \green}{\partial \tauy}\right) - \tauy^{{5/2}}\frac{\partial \green}{\partial \tauy} - \frac{\tauy^{{3/2}}\chi}{2}\frac{\partial \green}{\partial \chi} \\
& - \frac{N_0 \tauy_0^{2}\hat\kappa ^3\tauy^{2}\delta(\tauy-\tauy_0)\delta(\chi-\chi_0) \delta(q-q_0)}{4\pi x_0^2 R_g^3 (kT_e)^3 \hat v ^3} \ ,
\end{split}
\label{eq:nond}
\end{equation}
where the dimensionless time, $q$, is related to the time $t$ via
\begin{equation}
q \equiv \frac{\hat \kappa^{3/2}}{\hat v^{1/2}} \frac{t \, c}{R_g} \ ,
\label{eq:nondeq}
\end{equation}
and $q_0$, $\tauy_0$,  and $\chi_0$ denote the dimensionless injection time, injection radius, and injection energy, respectively, given by
\begin{equation}
q_0 = \frac{\hat \kappa^{3/2}}{\hat v^{1/2}} \frac{t_0 \, c}{R_g} \ , \ \ \
\tauy_0 = \frac{\hat v}{\hat \kappa}\frac{R_g}{r_0} \ , \ \ \
\chi_0 = \frac{\epsilon_0}{kT_e} \ .
\label{eq:nondim1}
\end{equation}
%


\section{Fourier Transform Method}
\label{sec:trans}

In order to compute theoretical time lags using Equation~(\ref{eq:lag}), we must use our transport equation formalism to evaluate the Fourier transforms of the light curves observed at the soft and hard channel energies, $\epsilon_s$ and $\epsilon_h$, respectively. The required solution for the Fourier transform can be obtained by performing a Fourier transformation of the partial differential transport equation, Equation~(\ref{eq:nond}). This yields an ordinary differential equation satisfied by the Green's function Fourier transform, $\Green$, which is related to the Green's function, $\green$, via the fundamental definition (cf. Equation~(\ref{eq:transforms}))
\begin{equation}
\Green(r,\epsilon,\omega) \equiv \int_{-\infty}^{+\infty} \green(r,\epsilon,t) \, e^{i \omega t} \, dt \ .
\label{eq:Fourier1}
\end{equation}
We can also express the Fourier transform in terms of the dimensionless time, $q$, by writing
\begin{equation}
\Green(r,\epsilon,\dimega) = \frac{\hat v^{1/2}}{\hat \kappa^{3/2}}
\frac{R_g}{c}\int_{-\infty}^{+\infty} \green(r,\epsilon,q) \, e^{i \dimega q} \, dq \ ,
\label{eq:Fourier2}
\end{equation}
where the dimensionless Fourier frequency, $\dimega$, is related to $\omega$ via
\begin{equation}
\dimega \equiv \frac{\hat v^{1/2} }{\hat \kappa^{3/2}} \frac{R_g}{c} \, \omega \ .
\label{eq:hatomega}
\end{equation}

Performing a Fourier transformation of Equation~(\ref{eq:nond}) by applying the operator $\int_{-\infty}^{+\infty} e^{i \dimega q} dq$ yields an ordinary differential equation for the Fourier transform Green's function, $\Green$. The result obtained is
\begin{equation}
\begin{split}
- i \dimega \Green = -\tauy^{{5/2}}\frac{\partial \Green}{\partial \tauy}
- \frac{\tauy^{{3/2}}\chi}{2}\frac{\partial \Green}{\partial \chi} \\
+ \tauy^{4} \frac{\partial}{\partial \tauy} \left(\tauy^{{-3/2}}  \frac{\partial \Green}{\partial \tauy} \right) \\
+ \frac{\Theta\,\tauy^{{3/2}}}{3\hat v \hat\kappa \chi^{2}}\frac{\partial}{\partial \chi}
\left[\chi^4 \! \left(\Green + \frac{\partial \Green}{\partial \chi}\right)\right] \\
+ \frac{N_0\tauy_0^{4}\hat\kappa^{3/2}\delta(\tauy-\tauy_0)\delta(\chi-\chi_0) e^{i \dimega q_0}}
{4\pi R_g^2 c (kT_e)^3 \chi_0^2 \hat v^{5/2}} .
\end{split}
\label{eq:FT}
\end{equation}
%


\subsection{Separation Functions}

In order to solve Equation~(\ref{eq:FT}) to determine the solution for the Fourier transform Green's function, $\Green$, we note that for $\chi \ne \chi_0$, the source term vanishes, and consequently Equation~(\ref{eq:FT}) can be separated using the function
\begin{equation}
F_\lambda(\tauy,\chi) = G(\tauy) \, H(\chi) \ ,
\label{eq:separate}
\end{equation}
where $\lambda$ denotes the separation constant, and the functions $G(\tauy)$ and $H(x)$ represent the spatial and energy separation functions, respectively. Substituting Equation~(\ref{eq:separate}) into Equation~(\ref{eq:FT}) and rearranging terms yields, we find that for $\chi \ne \chi_0$,
\begin{equation}
\begin{split}
- i \dimega \tauy^{{-3/2}} + \frac{\tauy}{G} \frac{dG}{d \tauy}
- \frac{\tauy^{{5/2}}}{G}\frac{d}{d\tauy}\left(\tauy^{{-3/2}}\frac{dG}{d\tauy}\right) \\
= \frac{\Theta}{3\hat v \hat\kappa}\frac{1}{\chi^2 H}\frac{d}{d\chi}\left[\chi^4 \! \left(H+\frac{dH}{d\chi}\right)\right]
- \frac{\chi}{2H} \frac{d H}{d \chi} = \frac{\lambda}{2} \ .
\label{eq:separation2}
\end{split}
\end{equation}

The separation constant $\lambda$ is independent of the coordinates $\tauy$ and $\chi$, and therefore we can obtain two distinct ordinary differential equations satisfied by the functions  $G(\tauy)$ and $H(\chi)$. The equations obtained are
\begin{equation}
\tauy \frac{d^2G}{d \tauy^{2}} - \left(\frac{3}{2} + \tauy\right)\frac{dG}{d\tauy}+\frac{\lambda}{2}\,G +  i \dimega \tauy^{{-3/2}} G = 0 \ ,
\label{eq:spatial}
\end{equation}
and
\begin{equation}
\frac{1}{\chi^2}\frac{d}{d\chi}\left[\chi^4 \! \left(H+\frac{dH}{d\chi}\right)\right]
- {\delta \chi} \frac{d H}{d \chi} - \delta \lambda H = 0 \ .
\label{eq:energy}
\end{equation}
where 
\begin{equation}
\delta \equiv \frac{3\hat v \hat\kappa}{2\Theta} \ .
\label{eq:delta}
\end{equation}
%


\subsection{Spatial Boundary Conditions}

The spatial boundary conditions utilized in this model are based on the transition to free-streaming that must occur at the inner and outer surfaces of the spherical coronal region. We note that the free-streaming or ``absorbing'' boundary condition was also discussed by Titarchuk et al. (1997). The specific (spatial) radiation flux, $\mathfrak{F}$, also referred to as the ``streaming function,'' is given by (Becker 1992)
\begin{equation}
\mathfrak{F}(\epsilon,r,t) = - \kappa \frac{\partial \green}{\partial r} - \frac{v_r\epsilon}{3}
\frac{\partial \green}{\partial \epsilon} \ .
\end{equation}
The inner and outer boundaries of the spherical coronal region are located at $r=r_{\rm in}$ and at $r=r_{\rm out}$, respectively, and the corresponding values of the dimensionless location parameter $y$ are (see Equation~(\ref{eq:traprad1}))
\begin{equation}
\begin{split}
y_{\rm in} &= \frac{R_g \hat v}{\hat \kappa \, r_{\rm in}} \ , \\
y_{\rm out} &= \frac{R_g \hat v}{\hat \kappa \, r_{\rm out}} \ .
\end{split}
\label{eq:yvals}
\end{equation}

The free-streaming boundary condition operative at the inner and outer boundaries can be written as
\begin{equation}
\begin{split}
- \kappa \frac{\partial \green}{\partial r} &= - c \green \ , \ \ r = r_{\rm in} \ ,  \quad {\rm (inner \ boundary)} \ , \\
- \kappa \frac{\partial \green}{\partial r} &= c \green \ , \quad r = r_{\rm out} \ ,  \quad {\rm (outer \ boundary)} \ .
\end{split}
\label{eq:free}
\end{equation}
Transforming the free-streaming boundary conditions from $r$ to $y$ yields
\begin{equation}
\begin{split}
 (\hat v\hat\kappa\tauy)^{1/2} \frac{\partial \green}{\partial \tauy} &= -  \, \green \ , \quad \tauy = \tauy_{\rm in} \ ,  \quad {\rm (inner \ boundary)} \ , \\
 (\hat v\hat\kappa\tauy)^{1/2} \frac{\partial \green}{\partial \tauy} &=   \, \green \ , \quad \tauy = \tauy_{\rm out} \ , \quad {\rm (outer \ boundary)} \ .
\end{split}
\label{fsbceq}
\end{equation}
Fourier transformation of Equations~(\ref{fsbceq}) demonstrates that the spatial boundary conditions satisfied by the Green's function Fourier transform, $\Green$, are given by
\begin{equation}
\begin{split}
 (\hat v \hat\kappa\tauy)^{1/2} \frac{\partial \Green}{\partial \tauy} &= -  \, \Green \ , \quad \tauy = \tauy_{\rm in} \ ,  \quad {\rm (inner \ boundary)} \ , \\
 (\hat v\hat\kappa\tauy)^{1/2} \frac{\partial \Green}{\partial \tauy} &=   \, \Green \ , \quad \tauy = \tauy_{\rm out}\ , \quad {\rm (outer \ boundary)} \ .
\end{split}
\label{fgbceq}
\end{equation}

By combining Equations~(\ref{eq:separate}) and (\ref{fgbceq}), we conclude that the spatial separation function $G(y)$ satisfies the free-streaming boundary conditions
\begin{equation}
 (\hat v\hat\kappa\tauy)^{1/2}G' + G = 0 \ , \qquad \tauy = \tauy_{\rm in} \ , \\
\label{eq:bcsIN}
\end{equation}
\begin{equation}
 (\hat v\hat\kappa\tauy)^{1/2}G' - G = 0  \ , \qquad \tauy = \tauy_{\rm out} \ ,
\label{eq:bcsOUT}
\end{equation}
where primes denote differentiation with respect to $\tauy$. The eigenfunctions, $G_n(\tauy)$, represent the discrete set of solutions to Equation~(\ref{eq:spatial}) that simultaneously satisfy both the inner and outer boundary conditions given by Equations~(\ref{eq:bcsIN}) and (\ref{eq:bcsOUT}), respectively. The corresponding eigenvalues for the separation constant $\lambda$ are denoted by $\lambda_n$. The solution procedure is further discussed below.

\subsection{Spatial Eigenfunctions}

The global solutions for the spatial eigenfunctions $G_n(\tauy)$ are obtained using a three-step process. The first step is to numerically solve the ordinary differential equation, Equation~(\ref{eq:spatial}), starting with the inner boundary condition (Equation~(\ref{eq:bcsIN})), imposed at $\tauy=\tauy_{\rm in}$. This yields the inner solution, denoted by $G_{\rm in}(\tauy)$. The second step is to carry out the numerical integration again, this time starting with the outer boundary condition (Equation~(\ref{eq:bcsOUT})), imposed at $\tauy=\tauy_{\rm out}$, which yields the outer solution, $G_{\rm out}(\tauy)$.

For general, arbitrary values of the separation constant, $\lambda$, the inner and outer solutions are linearly independent functions. However, for certain special values of $\lambda$, the Wronskian of the inner and outer solutions vanishes, i.e.,
\begin{equation}
\mathfrak{W}(\tauy_*) = 0 \ ,
\label{eq:wron1}
\end{equation}
where
\begin{equation}
\mathfrak{W}(\tauy_*) \equiv G_{\rm in}(\tauy_*) \, G'_{\rm out}(\tauy_*)
- G_{\rm out}(\tauy_*) \, G'_{\rm in}(\tauy_*) \ ,
\label{eq:wron2}
\end{equation}
and $\tauy_*$ is located anywhere in the computational domain, so that $\tauy_{\rm in} \le \tauy_* \le \tauy_{\rm out}$. The eigenvalues $\lambda_n$ are the roots of Equation~(\ref{eq:wron1}). We note that the Fourier frequency, $\dimega$, appears in Equation~(\ref{eq:spatial}), and therefore it follows that a unique set of eigenvalues $\lambda_n$ is obtained for each value of $\dimega$.


\subsection{Eigenfunction Orthogonality}

The final closed-form solution for the Fourier transform $\Green$ can be expressed as an infinite series if we can demonstrate that the spatial eigenfunctions, $G_n(\tauy)$, form an orthogonal set. We can establish the orthogonality of the spatial eigenfunctions as follows. First we rewrite Equation~(\ref{eq:spatial}) in the Sturm-Liouville form
\begin{equation}
\frac{d}{d\tauy} \left[S(\tauy)\frac{dG_n}{d\tauy}\right] + Q(\tauy) G_n + \frac{\lambda_n}{2} {\Omega}(\tauy) G_n = 0 \ ,
\label{eq:sturm}
\end{equation}
where the weight function, ${\Omega}(\tauy)$, is defined by
\begin{equation}
{\Omega}(\tauy) = \tauy^{{-5/2}}e^{-\tauy} \ ,
\end{equation}
and the functions $S(\tauy)$ and $Q(\tauy)$ are defined by
\begin{equation}
\begin{split}
S(\tauy) &= \tauy^{{-3/2}}e^{-\tauy} \ , \\
Q(\tauy) &= i \dimega \tauy^{{-4}}e^{-\tauy} \ .
\end{split}
\end{equation}

Next, we assume that $\lambda_n$ and $\lambda_m$ represent distinct eigenvalues associated with the spatial eigenfunctions $G_n$ and $G_m$, respectively. Equation~(\ref{eq:sturm}) is then multiplied by $G_m$ and the result is subtracted from the same equation with the indices $n$ and $m$ interchanged. After some algebra, the result obtained is
\begin{equation}
\begin{split}G_m\frac{d}{d\tauy}\left[S(\tauy)\frac{dG_n}{d\tauy}\right] - G_n\frac{d}{d\tauy}
\left[S(\tauy)\frac{dG_m}{d\tauy}\right] \\
= (\lambda_m - \lambda_n) {\Omega}(x) G_n G_m \ .
\end{split}
\end{equation}
Integration by parts between the inner and outer radii $\tauy_{\rm in}$ and $\tauy_{\rm out}$ yields
\begin{equation}
\begin{split}S(\tauy)\left[G_m\frac{dG_n}{d\tauy}-G_n\frac{dG_m}{d\tauy}\right]\bigg|_{\tauy_{\rm in}}^{\tauy_{\rm out}}
= \\ (\lambda_m - \lambda_n) \int_{\tauy_{\rm in}}^{\tauy_{\rm out}} {\Omega}(\tauy) G_n(\tauy) G_m(\tauy) \, d\tauy \ .
\end{split}
\label{SLOeq}
\end{equation}
Utilizing the spatial free-streaming boundary conditions at the inner and outer radii (Equations~(\ref{eq:bcsIN}) and (\ref{eq:bcsOUT})), we find that the left-hand side of Equation~(\ref{SLOeq}) vanishes. Hence we obtain
\begin{equation}
\int_{\tauy_{\rm in}}^{\tauy_{\rm out}} {\Omega}(\tauy) G_n(\tauy) G_m(\tauy) \, d\tauy = 0 \ , \quad \lambda_n \ne \lambda_m \ .
\label{SLO2eq}
\end{equation}
This establishes that the spatial eigenfunctions $G_n$ and $G_m$ are orthogonal with respect to the weight function ${\Omega}(\tauy)$ in the computational domain located between the inner and outer free-streaming boundaries located at $\tauy=\tauy_{\rm in}$ and $\tauy=\tauy_{\rm out}$, respectively.


\subsection{Energy Eigenfunctions}

The energy separation functions, $H(\chi)$, satisfy the second-order ordinary differential equation given by Equation~(\ref{eq:energy}). In addition, in order to obtain convergent results for the photon number and energy densities using Equations~(\ref{eq:nden}) and (\ref{eq:enden}), we note that $H$ must not increase faster than $\epsilon^{-3}$ as $\epsilon \to 0$. Likewise, $H$ must decrease faster than $\epsilon^{-4}$ as $\epsilon \to \infty$. Additionally, $H$ must be continuous at the injection energy $\chi=\chi_0$ in order to avoid an infinite diffusive flux in the energy space. Becker \& Wolff (2007) obtained the exact solution to Equation~(\ref{eq:energy}) satisfying the required boundary and continuity conditions. The result is given by their Equation~(48), which can be written as
\begin{equation}
H_n(\chi,\chi_0) = \chi^{\kappa-4}e^{-\chi/2}M_{\kappa,\mu}(\chi_{\rm min})W_{\kappa,\mu}(\chi_{\rm max}) \ ,
\label{eq:solutionenergy}
\end{equation}
where $M_{\kappa,\mu}$, and $W_{\kappa,\mu}$ denote the Whittaker functions,
\begin{equation}
\chi_{\rm min}\equiv {\rm min}(\chi,\chi_0) \ , \qquad \chi_{\rm max} \equiv {\rm max}(\chi,\chi_0) \ ,
\label{eq:ydef}
\end{equation}
and we have introduced the parameters
\begin{equation}
\kappa \equiv \frac{1}{2}\left(\delta + 4\right) \ , \qquad \mu \equiv \frac{1}{2}
\left[\left(3-\delta\right)^2 + 4 \, \delta \lambda_n\right]^{1/2} \ .
\label{eq:chidef}
\end{equation}
%


\subsection{Final Closed-Form Solution}

Having established that the spatial eigenfunctions form a complete orthogonal set, it is now possible to express the Fourier transform Green's function, $\Green$, using the infinite series
\begin{equation}
\Green(\tauy_0,\tauy,\chi_0,\chi,\dimega) = \sum_{n=1}^{N_{\rm max}} C_n G_n(\tauy, \tauy_0) H_n(\chi,\chi_0) \ ,
\label{eq:series}
\end{equation}
where $\tauy_0$ indicates the injection location, and $N_{\rm max}$ denotes the maximum index included in the sum, which must be large enough to ensure convergence of the numerical results for the time lags.
The computation of the time lags is further discussed in Section~\ref{sec:applications}.

The expansion coefficients, $C_n$, can be determined by exploiting the orthogonality of the spatial eigenfunctions and applying the derivative jump condition
\begin{equation}
\lim_{\varepsilon \to 0} \Delta\left[\chi^4\frac{\partial \Green}{\partial \chi}\right]\Bigg|_{\chi=\chi_0-\varepsilon}^{\chi=\chi_0+\varepsilon}= \frac{-\delta N_0\hat\kappa^{3/2}\delta(\tauy-\tauy_0)e^{i \dimega q_0}\tauy^{{5/2}}}{2\pi \hat v^{5/2}R_g^2 c (m_e c^2)^3 \Theta^3} \ ,
\label{eq:jump}
\end{equation}
which is obtained by integrating Equation~(\ref{eq:FT}) over a small region in energy space around $\chi=\chi_0$. Equation~(\ref{eq:series}) is now substituted into Equation~(\ref{eq:jump}) to obtain
\begin{equation}
\begin{split}
\sum_{n=1}^{N_{\rm max}} C_n G_n(\tauy) \chi_0^{\kappa-4}e^{-\chi_0/2} {\cal W}(\chi_0) \qquad \\
= \frac{-\delta N_0\hat\kappa^{3/2}\delta(\tauy-\tauy_0)e^{i \dimega q_0}\tauy^{{5/2}}}{2\pi \hat v^{5/2}R_g^2 c (m_e c^2)^3 \Theta^3 \chi_0^4} \ ,
\end{split}
\label{eq:almost}
\end{equation}
where the Wronskian, ${\cal W}(\chi_0)$, is defined by (Becker \& Wolff 2007)
\begin{equation}
\begin{split}
{\cal W}(\chi_0) &\equiv M_{\kappa,\mu}(\chi_0) W'_{\kappa,\mu}(\chi_0) - W_{\kappa,\mu}(\chi_0) M'_{\kappa,\mu}(\chi_0) \\
&= \frac{-\Gamma(1+2\mu)}{\Gamma(\mu-\kappa+1/2)} \ ,
\end{split}
\label{eq:wronse}
\end{equation}
and primes denote differentiation with respect to $\chi$. The final result in Equation~(\ref{eq:wronse}) is obtained using Equation~(55) from Becker \& Wolff (2007).

The solution for the expansion coefficients, $C_n$, is obtained by multiplying Equation~(\ref{eq:almost}) by $G_m(\tauy) {\Omega}(\tauy)$ and integrating with respect to $\tauy$ from $\tauy_{\rm in}$ to $\tauy_{\rm out}$. Utilizing the orthogonality of the spatial eigenfunctions (Equation~(\ref{SLO2eq})), after some algebra, the final expression obtained for $C_n$ is
\begin{equation}
C_n =  \frac{\delta N_0\hat\kappa^{3/2}G_n(\tauy_0)\Omega(\tauy_0)e^{\chi_0/2}e^{i \dimega q_0}\tauy_0{^{5/2}}\Gamma(\mu-\kappa+\frac{1}{2})}{2\pi \hat v^{5/2}R_g^2 c (m_e c^2)^3 \Theta^3 \chi_0^\kappa\mathfrak{I}_n\Gamma(1+2\mu)} \ ,
\label{eq:coeff}
\end{equation}
where the quadratic normalization integrals, $\mathfrak{I}_n$, are defined by
\begin{equation}
\mathfrak{I}_n \equiv \int_{\tauy_{\rm in}}^{\tauy_{\rm out}} {\Omega}(\tauy) \, G^2_n(\tauy) \, d\tauy \ .
\end{equation}

Combining Equations~(\ref{eq:series}) and (\ref{eq:coeff}), we find that the final closed-form solution for the Fourier transform Green's function, $\Green$, can be written as
\begin{equation}
\begin{split}
\Green (\tauy_0, \tauy,&\chi_0,\chi,\dimega)  = \\
& \frac{\delta N_0\hat\kappa^{3/2}\chi^{\kappa-4}\Omega(\tauy_0)e^{(\chi_0-\chi)/2}e^{i \dimega q_0}\tauy_0{^{5/2}}}{2\pi \hat v^{5/2}R_g^2 c (m_e c^2)^3 \Theta^3 \chi_0^\kappa} \hfil \\
&\times \sum_{n=1}^{N_{\rm max}} \frac{\Gamma(\mu-\kappa+\frac{1}{2})}{\mathfrak{I}_n \Gamma(1+2\mu)}
\, G_n(\tauy_0)G_n(\tauy) \\
&\times M_{\kappa,\mu}(\chi_{\rm min})W_{\kappa,\mu}(\chi_{\rm max}) \ ,
\end{split}
\label{eq:finalF}
\end{equation}
where $\chi_{\rm min}$, $\chi_{\rm max}$, $\kappa$, and $\mu$ are computed using Equations~(\ref{eq:ydef}) and (\ref{eq:chidef}). Equation~(\ref{eq:finalF}) gives the final solution for the Fourier transform, $\Green$, of the radiation Green's function, $\green$, inside the corona, evaluated at dimensionless energy $\chi$, dimensionless location $\tauy$, and dimensionless Fourier frequency $\dimega$, which correspond to the dimensional quantities $\epsilon$, $r$, and $\omega$, respectively. The solution describes the response to the impulsive injection of $N_0$ photons with energy $\epsilon_0$ at radius $r_0$ and time $t_0$, which correspond to the dimensionless location $\tauy_0$, dimensionless energy $\chi_0$, and dimensionless time $q_0$.

Based on the linearity of the fundamental transport equation (Equation~(\ref{eq:spheretransport})), it is straightforward to show that the Fourier transform associated with any desired photon source distribution, $Q$, can be evaluated using the integral convolution (cf. Equation~(\ref{eq:convolve}))
\begin{equation}
F(r,\epsilon,\omega) = \int_0^\infty \int_0^\infty \int_0^\infty \frac{\Green}{N_0} \, \epsilon_0^2 \, Q \, 4 \pi r_0^2 \, dr_0 \, d\epsilon_0 \, dt_0
\ .
\label{eq:convolve2}
\end{equation}
In practice, the upper and lower bounds for the radial integration in Equation~(\ref{eq:convolve2}) are replaced with the inner and outer radii of the quasi-spherical corona. In Section~\ref{sec:compute}, we use Equation~(\ref{eq:finalF}) to compute the Fourier transform of the {\it escaping} radiation field, which is the basis for the time lag calculations.


\section{Time Lag Computation}
\label{sec:compute}

The time-dependent X-ray photon count-rate spectrum observed at the detector, $\mathscr F_\epsilon(\epsilon)$, is simply related to the radiation distribution escaping from the outer edge of the spherical region, located at radius $r = r_{\rm out}$. Since we have employed a free-streaming boundary condition at the outer radius (Equation~(\ref{eq:free})), the computation of the escaping radiation spectrum is self-consistent, and we can therefore write the time-dependent X-ray photon count-rate spectrum observed at the detector as
\begin{equation}
\mathscr F_\epsilon(\epsilon,t) = c \left(\frac{r_{\rm out}}{D} \right)^2\epsilon^2 \green(r_{\rm out},\epsilon,t)
\ \ \propto~ \rm cm^{-2} \ s^{-1} \ erg^{-1} \ ,
\label{eq:theoryF}
\end{equation}
where $D$ is the distance to the source. In order to utilize the formalism developed here to compute the observed time lags, we need to evaluate the Fourier transform of the observed count-rate spectrum, denoted by $\tilde{\mathscr F}_\epsilon(\epsilon,\omega)$, given by
\begin{equation}
\tilde{\mathscr F}_\epsilon(\epsilon,\omega) = \int_{-\infty}^{+\infty}
\mathscr F_\epsilon(\epsilon,t) \, e^{i \omega t} \, dt \ ,
\label{eq:theoryF3}
\end{equation}
or, equivalently (cf. Equation~(\ref{eq:Fourier2})),
\begin{equation}
\tilde{\mathscr F}_\epsilon(\epsilon,\dimega) = \frac{\hat v^{1/2}}{\hat \kappa^{3/2}}\frac{R_g}{c}\int_{-\infty}^{+\infty}
\mathscr F_\epsilon(\epsilon,q) \, e^{i \dimega q} \, dq \ .
\label{eq:theoryF3b}
\end{equation}
Applying a Fourier transformation to both sides of Equation~(\ref{eq:theoryF}) by applying the operator $\int_{-\infty}^{+\infty} e^{i \dimega q} dq$ yields
\begin{equation}
\tilde{\mathscr F}_\epsilon(\epsilon,\dimega) = c \left(\frac{r_{\rm out}}{D} \right)^2\epsilon^2 \Green(r_{\rm out},\epsilon,\dimega) \ ,
\label{eq:theoryF2}
\end{equation}
where $F_{\rm G}$ is evaluated using Equation~(\ref{eq:finalF}).

The theoretical soft and hard energy light curves, $s(t)$ and $h(t)$, respectively, required to compute the time lags using Equation~(\ref{eq:lag}) are related to the photon count-rate spectrum, $\mathscr F_\epsilon(\epsilon,t)$, via
Equations~(\ref{eq:transforms2}). Applying a Fourier transformation to Equations~(\ref{eq:transforms2}) yields the corresponding expressions
\begin{equation}
\begin{split}
S(\tilde\omega) = \tilde{\mathscr F}_\epsilon(\epsilon_s,\tilde\omega) \ , \\
H(\tilde\omega) = \tilde{\mathscr F}_\epsilon(\epsilon_h,\tilde\omega) \ ,
\end{split}
\label{eq:theoryF5}
\end{equation}
where $S$ and $H$ denote the soft- and hard-energy light curve Fourier transforms, respectively, and the function $\tilde{\mathscr F}_\epsilon$ is evaluated using Equation~(\ref{eq:theoryF2}).

In order to compute X-ray time lags using our theoretical model, we must initially specify five parameters, namely: (1) the black hole mass, $M$; (2) the dimensionless accretion rate, $\dot m = \dot M/\dot M_{\rm E}$; (3) the dimensionless velocity parameter, $\hat v$; (4) the dimensionless inner radius, $\tilde r_{\rm in}=r_{\rm in}/R_g$; and (5) the dimensionless outer radius, $\tilde r_{\rm out}=r_{\rm out}/R_g$. We can then determine the value of $\hat \kappa$ using Equation~(\ref{eq:traprad2}), which gives $\hat \kappa = \hat v /(3 \dot m)$. This allows us to generate a corresponding vector of values for the dimensionless Fourier frequency, $\dimega$, using Equation~(\ref{eq:hatomega}).

After determining the set of eigenvalues and spatial eigenfunctions for each selected value of the dimensionless Fourier frequency, $\tilde \omega$, we must then vary the remaining theory parameters in order to obtain an acceptable qualitative fit to the observed time lags. The remaining four theory parameters comprise (1) the energy of the injected seed photons, $\epsilon_0$; (2) the dimensionless electron temperature $\Theta = k T_e/(m_e c^2)$; (3) the hard photon channel energy, $\epsilon_h$; and (4) the soft photon channel energy, $\epsilon_s$.


\section{Application to 1H~0707-495}
\label{sec:applications}

Fabian et al. (2009) and Zoghbi et al. (2010) have presented and discussed observations of time lags from the narrow-line Seyfert 1 galaxy 1H~0707-495. The time lags comprise an interesting patten, with the source displaying hard lags of magnitude $\sim 100\,$s for Fourier frequencies $\nu_F \lesssim 10^{-3}\,$Hz, and soft lags on the order of $\sim 10\,$s for Fourier frequencies $\nu_F \gtrsim 10^{-3}\,$Hz. We use our model to replicate the hard and soft lags observed from 1H~0707-495. The simulated time lags, $\delta t$, can be computed using Equation~(\ref{eq:lag}) once the Fourier-transformed soft and hard channel energy light curves, $S(\tilde\omega)$ and $H(\tilde\omega)$, respectively, have been evaluated using Equation~(\ref{eq:theoryF5}).

\subsection{Model Parameters}

The mass of the black hole at the center of 1H~0707-495 is estimated to be $M = 2 \times10^6\,\msun$ (Zhou \& Wang 2005). Assuming that the steady-state emission from the source is isotropic, the X-ray luminosity, $L_X$, is computed using
\begin{equation}
L_X = 4\pi D^2F_X \ ,
\label{eq:luminosity}
\end{equation}
where $F_X=1.44 \times 10^{-10} \rm \ erg\ cm^{-2}\ s^{-1}$ is the observed X-ray flux in the energy range 0.3-4\,keV, calculated using data from Zoghbi et al. (2010) and $D = 181$ Mpc is the luminosity distance (Sani et al. 2010). The X-ray luminosity obtained using Equation~(\ref{eq:luminosity}) is therefore $L_X = 5.63 \times 10^{44} \rm \ erg \ s^{-1}$.

We will assume here that the inner edge of the spherical region, at radius $r = r_{\rm in}$, is located at the radius of the innermost stable prograde circular orbit, $r_{\rm ISCO}$, so that
\begin{equation}
r_{\rm in} = r_{\rm ISCO} \ .
\label{eq:risco1}
\end{equation}
The value of $r_{\rm ISCO}$ is computed using (Shapiro \& Teukolsky 1983)
\begin{equation}
\begin{split}
&r_{\rm ISCO} = R_g\left[3 + Z_2 - \sqrt{\left(3 - Z_1\right)\left(3 + Z_1 + 2 Z_2\right)}\right] \ , \\
&Z_1 \equiv 1 + \left(1 - a^2\right)^{1/3} \left[\left(1 + a\right)^{1/3} + \left(1 - a\right)^{1/3}\right] \ , \\
&Z_2 \equiv \left(3 a^2 + Z_1^2\right)^{1/2} \ ,
\end{split}
\label{eq:risco}
\end{equation}
where $a = J/(McR_g)$ denotes the dimensionless spin parameter for a black hole with mass $M$ and angular momentum $J$. In the case of a non-rotating black hole, $a \to 0$, and Equation~(\ref{eq:risco}) reduces to the Schwarzschild result, $r_{\rm ISCO} = 6 R_g$. The radius of the event horizon for prograde orbits, $r_H$, is given by
\begin{equation}
r_H = R_g\left(1 + \sqrt{1-a^2}\right) \ ,
\label{eq:rH}
\end{equation}
which reduces to $r_H = 2 R_g$ in the Schwarzschild limit.

We set the spin parameter for the black hole using $a = 0.98$ (Fabian et al. 2009), in which case the inner edge of the spherical region is located at radius $r_{\rm in} = r_{\rm ISCO} = 1.61\,R_g$ (Equation~(\ref{eq:risco})), and the event horizon is located at $r_H = 1.20\,R_g$ (Equation~(\ref{eq:rH})). The outer edge of the spherical region is set at $r_{\rm out} = 120\,R_g$. Recall that in our model, the inner and outer radii are associated with the spherical corona, rather than the accretion disk (see Section 2). The value used for the dimensionless velocity parameter is $\hat v=\sqrt{2}$, which corresponds to exact free-fall.

We set the dimensionless accretion rate using $\dot m = \dot M/\dot M_{\rm E} = 1.1$, which is consistent with the findings of Tanaka et al. (2004), who estimated that $\dot M \gtrsim \dot M_{\rm E}$ in 1H~0707-495. According to Equation~(\ref{eq:traprad2}), this yields for the dimensionless diffusion parameter $\hat \kappa = 0.429$. The values of the inner and outer dimensionless location parameters, $y_{\rm in}$ and $y_{\rm out}$, can be computed using Equation~(\ref{eq:yvals}), which yield $y_{\rm in} = 2.04$ and $y_{\rm out}=0.028$. We note that according to Equation~(\ref{eq:traprad2}), the trapping radius is located at $r_t = 3.3 R_g$ for $\dot m =1.1$. The primary model parameters are listed in Table~1.


\begin{deluxetable}{cccccccc}[Ht]
\tablecaption{Model Parameters\label{tbl-1}}
\tablewidth{0pt}
\tablehead{
\colhead{$\dot m$~}
& \colhead{$\hat \kappa$}
& \colhead{$\hat v$}
& \colhead{$\Theta$}
& \colhead{$r_{\rm {in}}$}
& \colhead{$r_{\rm {out}}$}
& \colhead{$y_{\rm {in}}$}
& \colhead{$y_{\rm {out}}$}
}
\startdata
1.1
&0.429
&1.414
&0.05
&1.61\,$R_g$
&120\,$R_g$
&2.04
&0.028
\\
\enddata
\end{deluxetable}

\begin{figure}[h]
\centerline{\includegraphics[scale=1.1]{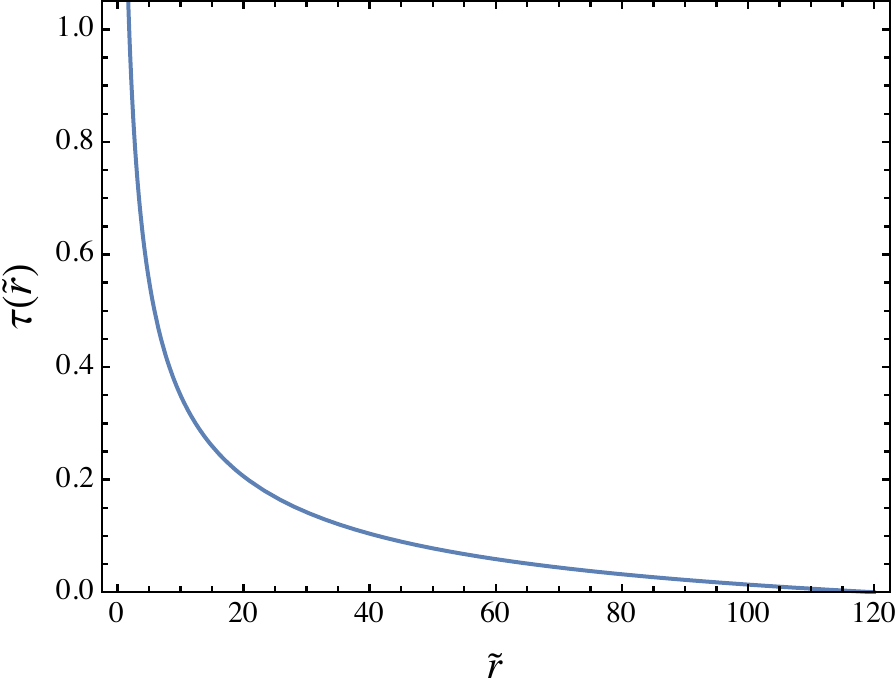}}
\caption{\footnotesize{The electron scattering optical depth, $\tau(\radw)$, for 1H~0707-495, measured inward from the outer edge of the spherical region, plotted as a function of the dimensionless radius $\radw$, computed using Equation~(\ref{eq:tau}).}}
\label{fig:tau}
\end{figure}

The electron scattering optical depth measured inward from the outer edge of the spherical region, $\tau(\radw)$, is computed using Equation~(\ref{eq:tau}), and plotted in Figure~\ref{fig:tau}. We note that the optical depth down to dimensionless radius $\radw = r/R_g = 1.86$ is close to unity, which is required in order to ensure that the relativistically blurred iron line emission near the black hole can be resolved spectroscopically, in reasonable agreement with the conclusions of (Wilkins et al. 2014). We find that $\tau = 1.08$ at $r=r_{\rm ISCO}$, and decreases monotonically out to $\tau = 0$ at $r=120\,R_g$.

\subsection{Simulated Time Lags}

Since the eigenvalues, $\lambda_n$, in our calculation are functions of the Fourier frequency, $\nu_F$, we must select a sample of discrete Fourier frequencies and evaluate the time lags at those frequencies. The reader should keep in mind that a continuous range of Fourier frequencies is not required in our model, since we do not need to perform the inverse Fourier transformation. Instead, the required time lags are computed directly from knowledge of the Fourier transform itself.

In our application to 1H~0707-495, we select values for the vector of Fourier frequencies, $\nu_F$ (in Hz), that are comparable to the values used by Fabian et al. (2009) and Zoghbi et al. (2010) in their plots of the time lags. The 12 values we adopt for the Fourier frequency are: $\nu_F= 7.94\times10^{-5}, 1.58\times 10^{-4}, 2.51\times 10^{-4}, 3.98\times 10^{-4}, 6.31\times 10^{-4}, 1.00\times 10^{-3}, 1.58\times 10^{-3}, 2.24\times 10^{-3}, 3.31\times 10^{-3}, 5.01\times 10^{-3}, 7.08\times 10^{-3}, 1.00\times 10^{-2}$. The 12 corresponding values for the dimensionless Fourier frequency, $\dimega$, computed using Equation~(\ref{eq:hatomega}) are: $\dimega$ = 0.021, 0.042, 0.066, 0.104, 0.166, 0.262, 0.416, 0.587, 0.869, 1.32, 1.86, 2.62. We note that the value of the photon injection number $N_0$ has no affect on the time lags, and we can set the injection time $t_0 = q_0 = 0$ without loss of generality.

\begin{figure}[h]
\centerline{\includegraphics[scale=0.65]{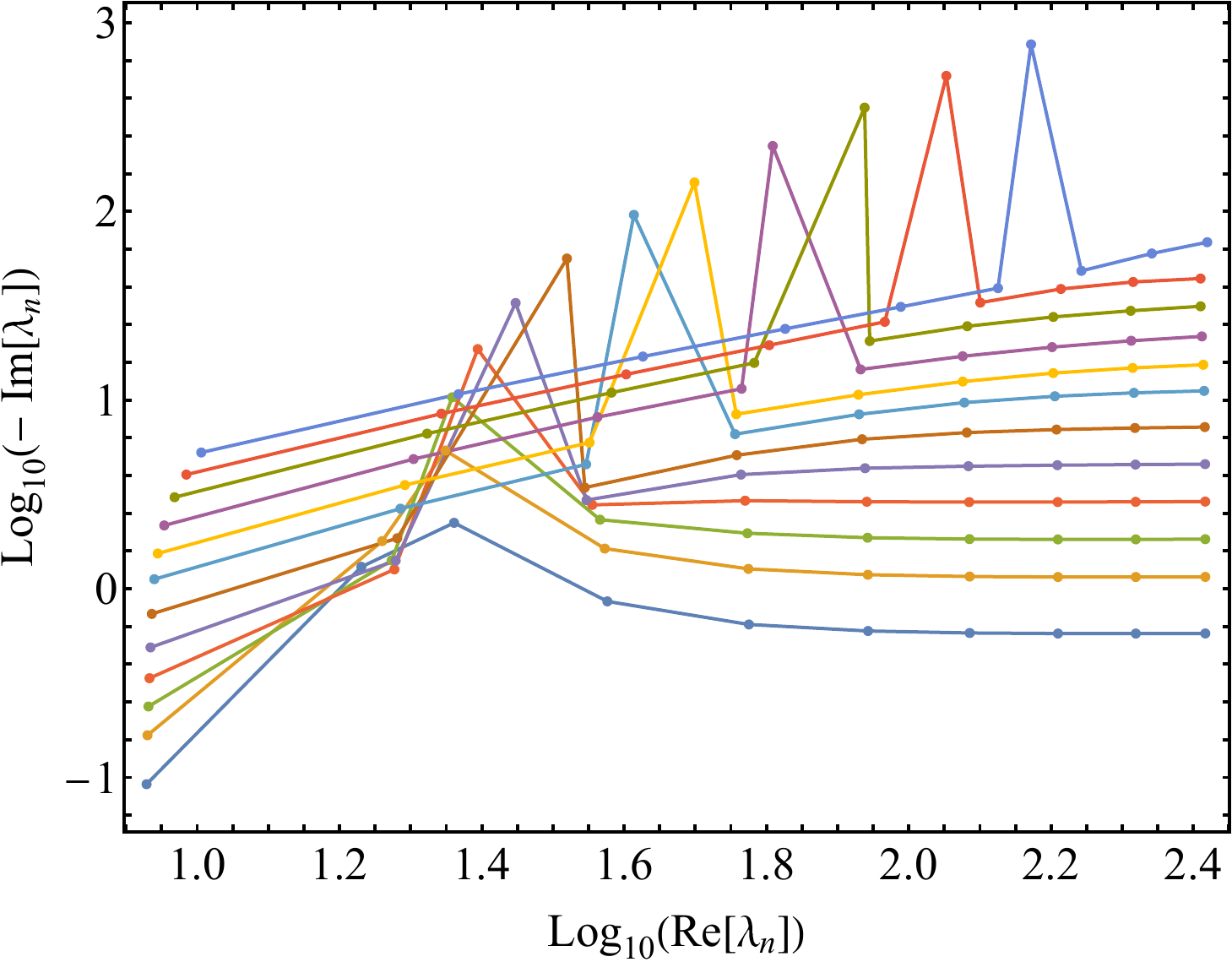}}
\caption{\footnotesize{Plot of the first 10 complex eigenvalues, $\lambda_n$, for each dimensionless Fourier frequency, $\dimega$, for 1H~0707-495, with the log of the real part on the horizontal axis and the log of the negative of the imaginary part on the vertical axis. The eigenvalue index $n$ increases from left to right, and $\dimega$ increases from bottom to top, through the vector of 12 values for $\dimega$. The eigenvalues are indicated by the points, and the colored lines are added for clarity, with each color corresponds to a different Fourier frequency. See the discussion in the text.}}
\label{fig:AGN2}
\end{figure}

The first 10 complex eigenvalues, $\lambda_n$, obtained for each discrete value of the Fourier frequency $\dimega$ adopted here are plotted in Figure~\ref{fig:AGN2}, with the value of $\dimega$ increasing from bottom to top. The eigenvalues are indicated by the points, and the colored lines are added to clarify the eigenvalue sequence for each frequency.

As noted in Section~\ref{sec:compute}, the computation of the simulated time lags, $\delta t$, using our theoretical model also requires specification of the injection radius $r_0$, the photon injection energy $\epsilon_0$, the electron temperature $\Theta = k T_e/(m_e c^2)$, and the hard and soft channel energies, $\epsilon_h$ and $\epsilon_s$, respectively. We set the value of the source energy using $\epsilon_0=0.89\,$keV, which notably correlates with the broadened and skewed energy of the iron L-line emission that Fabian et al. (2009) reported for 1H~0707-495. We set the value of the injection radius using $r_0=16R_g$, which yields $\tauy_0=0.206$ according to Equation~(\ref{eq:nondim1}). The value adopted for the dimensionless electron temperature is $\Theta = 0.05$, so that $k T_e = 25.6\,$keV. This yields for the electron temperature $T_e = 2.96 \times 10^8\,$K, which is consistent with the expected temperature range for the corona (e.g., Wilkins et al. 2014; Kara et al. 2017).

A remaining issue is how we set the soft and hard channel energies in our model, $\epsilon_s$ and $\epsilon_h$, respectively. In their time lag computations, Fabian et al. (2009) use $\epsilon_s = 0.3-1\,$keV for the soft energy band and $\epsilon_h = 1-4\,$keV for the hard energy band. However, our model is not able to accommodate an energy range for the soft and hard bands, and instead we need to specify exact values for $\epsilon_s$ and $\epsilon_h$. We set our hard and soft channel energies to the values $\epsilon_s = 0.33\,$keV and $\epsilon_h = 1.76\,$keV in order to approximate the variability found in these energy bands.

We have analyzed the convergence of the results obtained for the time lags as a function of the value of the maximum index, $N_{\rm max}$, appearing in the sum in Equation~(\ref{eq:finalF}). We find that setting $N_{\rm max} = 5$ is sufficient to ensure convergence to a relative error of $\sim 1\%$, and therefore we generally adopt the value $N_{\rm max} = 10$ as a conservative value when computing the time lags for each Fourier frequency $\omega$.

\begin{figure}[h]
\centerline{\includegraphics[scale=0.45]{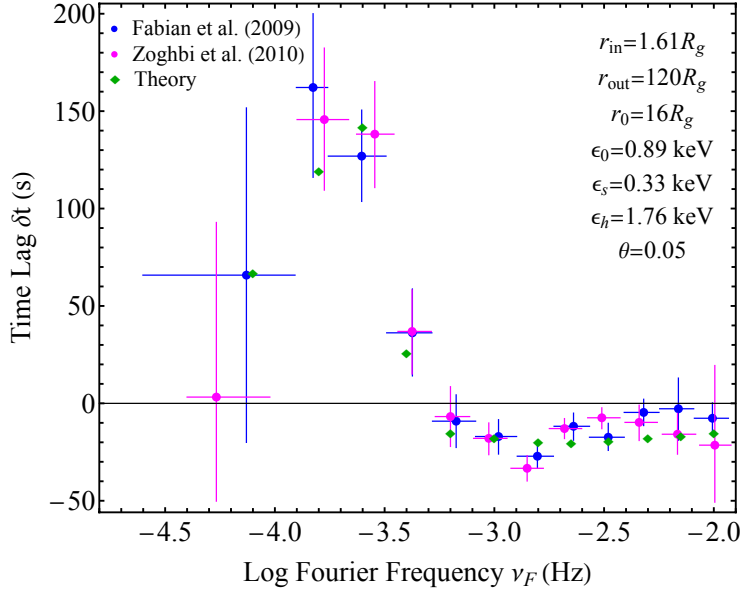}}
\vskip-0.9truein
\caption{\footnotesize{Time lag distribution $\delta t$ for 1H~0707-495, computed using Equation~(\ref{eq:lag}) plotted as a function of the Fourier frequency, $\nu_F = \omega/(2 \pi)$ (green diamonds). According to our sign convention, a positive value for $\delta t$ indicates a hard lag and a negative value indicates a soft lag. The observational data are taken from Fabian et al. (2009; blue dots) and Zoghbi et al. (2010; purple dots).}}
\label{fig:AGN3}
\end{figure}

The resulting time lag distribution, $\delta t$, computed using Equation~(\ref{eq:lag}), is plotted as a function of the Fourier frequency $\nu_F = \omega/(2 \pi)$ in Figure \ref{fig:AGN3}, and compared with the observational data for 1H~0707-495 reported by Fabian et al. (2009) and Zoghbi et al. (2010). Note the good qualitative agreement between the theoretical model and the time-lag data from Fabian et al. (2009) and Zoghbi et al. (2010), including the change in sign of $\delta t$ from positive (hard lag) below Fourier frequency $\sim 5 \times 10^{-4}\,$Hz to negative (soft lag) at higher frequencies.

%


\section{DISCUSSION AND CONCLUSION}
\label{sec:conclusions}

The short time lags of $\sim 10-100$~seconds detected from many AGNs imply that the physical processes are occurring in the inner region of the accretion flow onto the black hole. Hence, the analysis and interpretation of the time lags provides a unique opportunity to probe the detailed structure of the innermost portion of the accretion flow, including the temperature and density profile, as well as the accretion morphology.

This has motivated us to develop a new model for the production of the observed hard and soft time lags from AGNs based on a combination of thermal and bulk Comptonization occurring in the quasi-spherical inner region of the accretion flow onto the black hole.
In our model, the time-dependent emission that is responsible for generating the time lags is produced in the quasi-spherical region, and the steady-state emission, comprising the bulk of the X-ray signal, is produced in the surrounding, geometrically thin but optically thick accretion disk (Shakura \& Sunyaev 1973).

We have applied our new model to the interpretation of the X-ray time lags observed from 1H~0707-495, as reported by Fabian et al. (2009) and Zoghbi et al. (2010). The theoretical results for the time lags plotted in Figure~\ref{fig:AGN3} indicate good qualitative agreement with the time lag data for 1H~0707-495, suggesting that the observed soft lags may be produced via thermal and bulk Comptonization occurring within the quasi-spherical region of the accretion flow. Hence in our model, the hard and soft time lags are both produced in a single region, via the action of a unified physical mechanism. 

Our physical model is based on a rigorous theoretical framework describing the reprocessing of an instantaneous flash of monochromatic photons injected at radius $r_0 = 16\,R_g$. The value for the injection energy adopted here, $\epsilon_0=0.89\,$keV, is approximately at the peak value for the energy of the broad iron L-line reported by Fabian et al. (2009), and hence may be due to fluorescence of the iron line, perhaps driven by inhomogeneities or ``clumps'' in the accretion flow (e.g., Merloni et al. 2006; Guti\'errez et al. 2021).

In the reverberation interpretation, the broadened iron L-line feature observed in the steady-state X-ray spectrum from 1H 0707-495 is a spatial reflection feature (e.g., Zoghbi et al. 2010). This is similar to our model, except that in our model, the seed photons may correlate with broadened iron L-line emission generated inside the corona. Hence the location of the source for the L-line emission in the steady-state spectrum is not necessarily identical with the source location for the impulsive injection of the seed photons used in our calculation of the time lags, although the physical emission mechanism may be the same.

In our application to 1H 0707-495, we have approximated the broadened iron L-line emission using a monochromatic source. This approximation can be generalized in future work using an integral convolution of the Green's function (Equation (\ref{eq:convolve2})) with the broadened source distribution. However, we do not expect this modification to substantially alter the results presented here.

The size of the corona assumed here bears further discussion. Morgan et al. (2012) published microlensing results that constrain the X-ray corona of lensed quasars to be smaller than $\lesssim 10 \, R_g$, which would seem to be a potential conflict with the size of the corona in our model, which is $r_{\rm out} \sim 120 \, R_g$. This problem can be resolved by noting that the overall scattering optical depth of the corona in our model is $\tau \sim 1.08$, and therefore the distribution of the last scattering over space indicates that many photons will escape promptly from deep layers without much scattering. Hence it follows that the apparent size of the X-ray emitting region in our model is not necessarily equal to the size of the corona implied by the microlensing observations.

We assume that the flash of seed photons is produced simultaneously at all points on a spherical shell with radius $r_0$. Of course, relativity precludes any coherence in the emission process occurring on timescales shorter than the light-crossing time for the emitting region. In this regard, it's important to note that one would obtain the same result for the Fourier transform solution that we obtain here even if the iron L-line photons where emitted at random locations on the spherical surface. This is because the time window for the Fourier transform ``sweeps up'' many such emission episodes. Hence, provided they are distributed in a spherically symmetric way, the total Fourier transform would remain in agreement with that computed here, and there would be no causality violation.

The simplicity of our model is in marked contrast to previous models, which generally rely on unusual geometrical configurations, or multiple thermal zones with different properties, in order to reproduce the observed time lags. For instance, Zoghbi et al. (2010) and Miller et al. (2010) invoke different physical mechanisms to explain the production of the hard time lags at small Fourier frequencies and the soft time lags at large Fourier frequencies. Conversely, in our model, both the hard and soft lags are produced as a natural consequence of thermal and bulk Comptonization occurring in the quasi-spherical inner region of the accretion flow.

It is interesting to further explore the behavior of the time lag $\delta t$ as a function of the Fourier frequency $\nu_F = \omega/(2 \pi)$. In particular, we note the change in sign of the time lags, at what we term the ``critical frequency,'' $\nu_c \sim 5 \times 10^{-4}\,$Hz (see Figure~\ref{fig:AGN3}). The complex behavior of $\delta t$ as a function of Fourier frequency, and the sign change between hard and soft lags at the critical frequency, stem from the fact that the energy of the injected iron L-line emission, $\epsilon_0 = 0.89\,$keV, lies between the soft and hard detector band energy ranges used to compute the time lags. The existence of hard lags for $\nu \lesssim \nu_c$ reflects the fact that on long time scales, recoil losses are more rapid than stochastic energization. Hence the drift towards high energies is slower than the drift towards lower energies, and this naturally produces hard time lags. Conversely, the appearance of soft lags for $\nu \gtrsim \nu_c$ reflect the rapid energization of photons that occurs in a single collision with an electron, which overwhelms recoil losses on short time scales (Lewis et al. 2016).

De Marco et al. (2013) noted that the Fourier frequency associated with the onset of negative lags (i.e., soft lags), denoted by $\nu_{\rm neg}$, scales as $\nu_{\rm neg} \propto (\dot M / \dot M_{\rm E})/(M)$. For sources accreting near the Eddington limit, with $\dot M \sim \dot M_{\rm E} \propto M$, the negative lag frequency scales as $\nu_{\rm neg} \propto M^{-1}$. Within the context of our model for bulk and thermal Comptonization, the negative lag frequency of De Marco et al. (2013) is roughly equivalent to our ``critical frequency,'' $\nu_c$, at which the time lags transition from soft to hard. We therefore find that the universality of the mass scaling law discovered by De Marco et al. (2013) can then be understood as a natural consequence of our model, since, according to Equation~(\ref{eq:hatomega}),
\begin{equation}
\omega_c = \dimega_c \left(\frac{\hat v^{1/2} }{\hat \kappa^{3/2}} \frac{R_g}{c}\right)^{-1} \, \ ,
\label{eq:hatomega2}
\end{equation}
where $\omega_c = 2 \pi \nu_c$ denotes the critical Fourier frequency, and the dimensionless critical frequency $\tilde \omega_c = 0.131$ is a constant in our model. Hence Equation~(\ref{eq:hatomega2}) implies that the critical frequency $\omega_c$ is inversely proportional to the mass of the black hole, in agreement with the results of De Marco et al. (2013). This strongly suggests that thermal and bulk Comptonization play an important role in spectral formation in all AGNs displaying soft time lags, regardless of mass scale.

In this paper we have presented a new model for the production of both hard and soft time lags in AGN via the thermal and bulk Comptonization of seed photons, which may be injected via fluorescence of the iron L-line. We developed a specific detailed calculation applicable to 1H 0707-495 and demonstrated good agreement with the time lag observations for this source. Although our model does not include an explicit reflection component, nonetheless we believe it is complementary to the previous models because it includes reverberation in the energy space via bulk and thermal Comptonization, as well as reverberation in the physical space, due to photon diffusion through the corona. In this sense, the model explored here explains all of the time lags observations for 1H 0707-495 using a single unified physical mechanism.

We note that in other sources there are soft lag features observed at higher energies than those observed from 1H 0707-495, which may be related to iron K-line emission. In future work we plan to extend our model to treat additional photon sources covering a broader energy range in order to more fully explore the role of thermal and bulk Comptonization in producing both hard and soft time lags from AGNs.

Although the effects of general relativity have not been fully incorporated here, we expect that the same fundamental results would be obtained in an equivalent, fully relativistic model, because the fundamental physical process leading to the production of the time lags is electron scattering in the relatively cool, spherical region of the accretion flow, which is rigorously treated in our model.

We are grateful to the anonymous referee for providing numerous suggestions for clarification that led to substantial improvements in the manuscript.


\end{document}